\documentclass[sigconf]{acmart}

\renewcommand\footnotetextcopyrightpermission[1]{} 
\usepackage[ruled]{algorithm2e} 

\newcommand{\slim}{{\scshape Slim}}
\newcommand{\wmf}{{\scshape wmf}}
\newcommand{\cdae}{{\scshape cdae}}

\newcommand{\mvae}{${\textsc{Mult-vae}}^{\textsc{ pr}}$}
\newcommand{\mdae}{{\scshape Mult-dae}}
\newcommand{\sae}{${\textsc{ease}}^{\textsc{r}}$}

\newcommand{\W}{B}
\newcommand{\X}{X}
\newcommand{\Y}{X}
\newcommand{\RR}{\mathbb{R}}
\newcommand{\EE}{\mathbb{E}}
\newcommand{\I}{\mathcal{ I}}
\newcommand{\UU}{\mathcal{U}}
\newcommand{\Wzero}{\hat{\W}  }
\newcommand{\PP}{\hat{P}}
\newcommand{\diag}{{\rm diag}}
\newcommand{\dMat}{{\rm diagMat}}
\newcommand{\N}{\mathcal{N}}

\begin{document}

\title{Embarrassingly Shallow Autoencoders for Sparse Data}
\titlenote{This paper is published in the proceedings of 'The Web Conference' (WWW) 2019,
  under the Creative Commons Attribution 4.0 International (CC-BY 4.0) license.}

\author{Harald Steck}
\affiliation{%
  \institution{Netflix}
  \city{Los Gatos}
  \state{California}
}
\email{hsteck@netflix.com}

\begin{abstract}
Combining simple elements from the literature, we define a linear model that is geared toward  sparse data, in particular implicit feedback data for recommender systems. We show that its training objective has a closed-form solution, and discuss the resulting conceptual insights.  Surprisingly, this simple  model  achieves better ranking accuracy than  various state-of-the-art collaborative-filtering approaches, including deep non-linear models,  on most of the publicly available data-sets used in our experiments. 
\end{abstract}


\maketitle

\section{Introduction}
Many recent improvements in  collaborative filtering can be attributed to deep learning approaches, e.g, \cite{hidasi15,hidasi17,liang18,sedhain15, zheng16,wu16,he17,cheng16}. Unlike in areas like  computer vision, however, it was found that a  \emph{small} number of hidden layers achieved the best recommendation accuracy. In this paper, we take this to the extreme, and define a linear model \emph{without}  a hidden layer (see Figure \ref{fig_sae}). The (binary) input vector indicates which items a user has interacted with, and the model's objective (in its output layer) is to predict  the best items to recommend to the user. This is done by reproducing the input as its output, as is typical for  \emph{auto}encoders.\footnote{Note, however, that the proposed model  does not follow the typical architecture of autoencoders, being comprised of an encoder and a decoder: one may introduce an implicit  hidden layer, however, by a (full-rank) decomposition of the learned weight-matrix of this model.}
 We hence  named it Embarrassingly Shallow AutoEncoder (in Reverse order: \sae{}).

This paper is organized as follows: we define \sae{}  in the next section, using simple elements from the literature. In Section \ref{sec_closedform}, we derive the closed-form solution of its convex training objective. This has several implications:  (1) it reveals that the neighborhood-based approaches used in collaborative filtering are based on conceptually incorrect item-item similarity-matrices, while \sae{}  may be considered  a principled neighborhood model, see Sections \ref{sec_interpretation} and \ref{sec_nn};
(2) the code for training \sae{}  is comprised of only a few lines, see Section \ref{sec_algo} and Algorithm 1; (3) if the model fits into memory, the wall-clock time for training \sae{}  can be   several  orders of magnitude less than  for training a \slim{}  \cite{ning11} model (see Section \ref{sec_complexity}), which is the most similar model to \sae{}.
Apart from that, we surprisingly found that \sae{}   achieved competitive ranking accuracy, and even outperformed  various deep, non-linear, or probabilistic models as well as neighborhood-based approaches on most of the publicly available data-sets used in our experiments (see Section \ref{sec_exp}).

\section{Model Definition}
Like in many recommender papers that use implicit feedback data,
we assume that the  data are given in terms of a sparse (typically binary\footnote{Non-binary matrices may also be used.})  matrix $\X \in \RR^{|\UU|\times|\I|}$, regarding the sets of users $\UU$ and items $\I$, where $|\cdot|$ denotes the size of a set. A positive value (typically 1) in $\X$ indicates that the user interacted with an item, while a value of 0 indicates that no interaction has been observed.

\begin{figure}[t]
\begin{center}
\includegraphics[height=2cm]{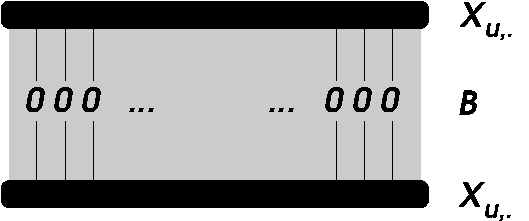}
\end{center}
\caption{The self-similarity of each item is constrained  to zero between the input and output layers.}
\label{fig_sae}
\end{figure}

The parameters of the \sae{}  model are given by the item-item  weight-matrix $\W \in \RR^{|\I|\times|\I|}$. Note that this is also similar to neigh-borhood-based approaches, see Section \ref{sec_nn}. In this weight matrix, self-similarity of an item in the input-layer with itself in the output layer is forbidden as to force the model to generalize when reproducing  the input as its output (see Figure \ref{fig_sae}): 
hence   the  diagonal of the weight-matrix is constrained to zero, $\diag(\W)=0$. This constraint is crucial, and is discussed in detail in the remainder of this paper. This constraint was first introduced in the \slim{} model  \cite{ning11}.

The predicted score   $S_{u,j}$ for an item $j\in \I$ given a user  $u\in\UU$ is defined by the dot product
\begin{equation}
S_{uj}=\X_{u, \cdot} \cdot \W_{\cdot,j},
\label{eq_score}
\end{equation}
where $\X_{u, \cdot}$ refers to  row $u$, and $\W_{\cdot,j}$  to column $j$. 
\section{Model Training}
We use the following convex objective for learning the weights $\W$:
\begin{eqnarray}
\min_\W &||\Y-\X\W||_F^2 + \lambda\cdot ||\W||_F^2 \label{eq_opt}\\
{\rm s.t.}& \diag( \W)=0 \label{eq_const}
\end{eqnarray}
Several comments are in order:
\begin{itemize}
\item We choose the square loss ($||\cdot||_F$ denotes the Frobenius norm) between the data $\Y$ and the predicted scores $S=X\W$  over other loss functions because it allows for a closed-form solution (see next section). Training with other loss functions, however, might result in improved ranking-accuracy: in \cite{liang18}, it was observed that  the multinomial likelihood resulted in better ranking-accuracy than training with the logistic likelihood (log loss) or the Gaussian likelihood (square loss).  Directly optimizing a (surrogate) ranking loss might result in further accuracy gains--however, at possibly increased computational costs.
\item We use L2-norm regularization of the weights $\W$ to be learned. The training objective hence has a single hyperparameter $\lambda$, to be optimized on a  validation set.

\item The constraint of a zero diagonal, $\diag( \W)=0$,  is crucial as to avoid the trivial solution $\W=I$ (self-similarity of items), where $I$ is the identity matrix. It was introduced in \slim{} \cite{ning11}.
\end{itemize}

\subsection{Closed-Form Solution}
\label{sec_closedform}
In this section, we show that the constrained convex optimization problem for learning the weight matrix $\W$ in Eqs. \ref{eq_opt} and \ref{eq_const} can be solved in closed form. 

We start by including the equality constraint in Eq. \ref{eq_const} into the objective function in Eq. \ref{eq_opt} by forming the Lagrangian
\begin{equation}
L=||\Y-\X\W||_F^2 + \lambda\cdot||\W||_F^2 + 2\cdot\gamma^\top\cdot \diag(\W),
\nonumber
\end{equation}
where  $\gamma=(\gamma_1,...,\gamma_{|\I|})^\top$ is the vector of Lagrangian multipliers. Its values  will be chosen in  Eq. \ref{eq_lag} such that the constraint in Eq. \ref{eq_const} is fulfilled.

The constrained optimization problem in  Eqs. \ref{eq_opt} and \ref{eq_const} is  solved by minimizing this Lagrangian. As a necessary condition, we hence set its derivative  to zero, which yields the estimate $\Wzero$ of the weight matrix  after re-arranging terms:
\begin{equation}
\Wzero = (\X^\top \X + \lambda I)^{-1} \cdot(\X^\top \Y - \dMat(\gamma)  ) ,
\nonumber
\end{equation}
where $\dMat(\cdot)$ denotes the  diagonal matrix and $I$ the identity matrix.
Defining (for sufficiently large $\lambda$)
\begin{equation}
\PP  \triangleq  (\X^\top \X + \lambda I)^{-1},
\label{eq_pp}
\end{equation}
this can be substituted into the previous equation:
\begin{eqnarray}
\Wzero &=& (\X^\top \X + \lambda I)^{-1} \cdot(\X^\top \Y - \dMat(\gamma)  )\nonumber\\
 &=& \PP \cdot(\PP^{-1} - \lambda I - \dMat(\gamma)  )\nonumber\\
 &=& I- \PP \cdot \left( \lambda I + \dMat(\gamma)  \right)\nonumber\\
 &=& I- \PP \cdot  \dMat(\tilde{\gamma})  
\label{eq_11}
\end{eqnarray}
where we defined the vector $\tilde{\gamma} \triangleq \lambda \vec{1}+\gamma $ in the last line, with $\vec{1}$ denoting a vector of ones.
The values of the Lagrangian multipliers $\gamma$, and hence $\tilde{\gamma}$, are determined by  the constraint $\diag(\Wzero)=0$. It follows  from Eq. \ref{eq_11}:
\begin{equation}
0=\diag(\Wzero) = \vec{1}  - \diag(\PP) \odot \tilde{\gamma}
\label{eq_lag}
\end{equation}
where $\odot$ denotes the elementwise product, and hence:
\begin{equation}
\tilde{\gamma} = \vec{1} \oslash \diag(\PP) , \nonumber
\end{equation}
where $\oslash$ denotes the elementwise division (which is well-defined given that $\PP$ is invertible).
Substituting this  into Eq. \ref{eq_11} immediately results in the closed-form solution:
\begin{equation}
\Wzero =   I- \PP \cdot \dMat\left( \vec{1}\oslash \diag(\PP)\right) .
\label{eq_w1}
\end{equation}
In other words, the learned weights  are given by:
\begin{equation}
\Wzero_{i,j} = \left\{
  \begin{array}{ll}
     0& {\rm if}\,\,\, i=j\\
     -\frac{\PP_{ij}}{\PP_{jj}}&{\rm otherwise.}
  \end{array}
\right .
\label{eq_w2}
\end{equation}
This solution obviously obeys the constraint of a zero diagonal. The off-diagonal elements are determined by the matrix $\PP$ (see Eq. \ref{eq_pp}), where the $j^{\rm th}$ column is divided by its diagonal element $\PP_{jj}$. Note that $\Wzero$ is an asymmetric matrix in general, while $\PP$ is symmetric (see Eq. \ref{eq_pp}).

Eqs. \ref{eq_pp} and \ref{eq_w2} show that the sufficient statistics for estimating $\W$ is given by the data Gram-matrix $G \triangleq \X^\top \X$, which is an item-item matrix. This is a consequence of using the square loss in Eq. \ref{eq_opt}, and is helpful for  estimating $\W$ from sparse data $\X$: if $\X$ is a sparse binary matrix, then $G=\X^\top \X$ is a co-occurrence matrix. The uncertainty of a co-occurrence count $G_{ij}$ is (approximately) determined by the standard deviation of the Poisson distribution, which is $\sqrt{G_{ij}}$. As long as the co-occurrence counts $G_{ij}$ are 'sufficiently large', $G$ and hence $\W$ can be estimated with small error. An interesting fact is that the entries of $G=\X^\top \X$ can be increased by two different mechanisms: (1) a denser $\X$ (due to users with increased activity), or (2) an increased number of  users in $\X$. The latter is particularly useful, as an increased sparsity of $\X$ can be compensated by an increased number of users. In other words, the problem that there possibly is only a small amount of data available for \emph{each} user (i.e., data sparsity), does not affect the uncertainty in  estimating  $\W$ if the number of users in the data matrix $\X$ is sufficiently large.   
\subsection{Interpretation}
\label{sec_interpretation}
In this section, we outline that the closed-form solution in Eq. \ref{eq_w2} does not come as  a complete surprise. To this end, let us consider the following special case throughout this section: let  the training data $\X$ be an i.i.d. sample of $|\UU|$ data points regarding a vector of $|\I|$ random variables $x\sim \N(0,\Sigma)$ that follows a Gaussian distribution with zero mean and covariance matrix $\Sigma \in \RR^{|\I|\times |\I|}$.

Then  the  estimate of the covariance matrix is $\hat{\Sigma}=\X^\top \X / |\UU|$. If we further drop the L2-norm regularization in Eq. \ref{eq_pp} and assume  invertibility, then  $\PP = \hat{\Sigma}^{-1}$ is the estimate of the so-called precision (or concentration) matrix. 
 Estimating the precision matrix from given data is a  main objective in  the area of graphical models (e.g., see \cite{lauritzenbuch96}).

It is well known \cite{lauritzenbuch96,besag75} that the (univariate) \emph{conditional distribution} of the random variable $x_j$ given the vector of \emph{all} the other variables, denoted by  $x_{-j} \triangleq (x_k)_{k \in \I\setminus \{j\}}$, is  a Gaussian distribution with variance 
${\rm var}(x_j | x_{-j})=1/P_{j,j}$   
and mean 
\begin{eqnarray}
\mu_{j|-j} \triangleq  \EE[x_j|x_{-j}]&= &
- x_{-j}\cdot P_{-j,j} / P_{j,j}\nonumber\\
&=& x_{-j}\cdot \W_{-j,j} \nonumber\\
&=& x\cdot \W_{.,j}\nonumber
\end{eqnarray}
where the dot in the first line denotes the dot-product between the (row) vector $x_{-j}$ and the $j^{\rm th}$ column of the precision matrix $P=\Sigma^{-1}$, omitting the  $j^{\rm th}$ element.  The second line follows from Eq. \ref{eq_w2}, and the last line from  $\W_{jj}=0$.
Note that this is identical to the prediction rule of the \sae{}  model  in Eq. \ref{eq_score}. This shows that \sae{}  makes a principled point-prediction that  user $u$ will like item $j$ \emph{conditional} on the fact that the user's past interactions with \emph{all}   items are given by $\X_{u,.}=x$.

A more well-known fact   (e.g., see \cite{meinshausen06}) is that the absence of an edge between the random variables $x_i$ and $x_j$ in a Markov network corresponds to the \emph{conditional independence} of the random variables $x_i$ and $x_j$ given the vector of \emph{all} the other variables $(x_k)_{k \in \I\setminus \{i,j\}}$, which is also equivalent to a zero entry in the precision matrix. Note that this is different from a zero entry in the covariance matrix $\Sigma$, which signifies  \emph{marginal independence} of $x_i$ and $x_j$. This shows that the precision matrix is the conceptually correct similarity-matrix to be used, rather than the covariance matrix, which (or some rescaled variant thereof) is typically used in state-of-the-art  neighborhood-based approaches (see Section \ref{sec_nn}).

Learning the graph structure in Markov networks  corresponds to learning a sparse precision matrix from data. Approaches developed in that field (e.g., see \cite{schmidtthesis} and references therein) might be useful for improved learning of a \emph{sparse} matrix $\Wzero$. This is beyond the scope of this paper. 

While the interpretation we outlined in this section is limited to the special case of normally distributed 
variables with \emph{zero mean}, note that the derivation of Eq. \ref{eq_w2} does not require that each column of the data matrix $\X$ has zero mean. In other words, in \sae{}  any definition of  the Gram matrix $G \triangleq \X^\top \X$ may be used, e.g., $G$ may be the co-occurrence matrix (if $\X$ is a binary matrix), proportional to the covariance matrix (if $\X$ is pre-processed to have zero mean in each column), or the correlation matrix (if each column of $\X$ is pre-processed to have zero mean and unit variance). After training \sae{}  on  these transformed matrices $\X$ and then transforming the predicted scores back to the original space (as defined by the binary matrix $\X$), we found in our experiments that the differences in the obtained ranking accuracies were of the order of the standard error, and we hence do not separately report these results in Section \ref{sec_exp}.

\subsection{Algorithm}
\label{sec_algo}
The Python code of the resulting learning algorithm is given  in Algorithm \ref{algo}. 
Note that the training of \sae{}  requires only the item-item matrix  $G=\X^\top \X$ as input, instead of the user-item matrix $\X$, and  hence is particularly efficient if the size of $G$ (i.e., $|\I|\times|\I|$) is smaller than the number of user-item-interactions in $\X$. In this case, the expensive computation of $G=\X^\top \X$ can be done on a big-data pre-processing system, prior to the actual model training.

\subsection{Computational Cost}
\label{sec_complexity}
The computational complexity of Algorithm \ref{algo} is determined by the matrix inversion of the data Gram-matrix $G \triangleq \X^\top X \in  \mathbb{R}^{|\I|\times|\I|}$ , which is $\mathcal{O}(|\I|^{3})$ when using a basic approach, and about $\mathcal{O}(|\I|^{2.376})$ when using the Coppersmith-Winograd algorithm. Note that this is independent of the number of users as well as the number of user-item-interactions, as $G$ can be computed in the pre-processing step. 

This computational complexity is orders of magnitude lower than the cost of training a \slim{} model and its variants \cite{ning11,levy13,sedhain16}: those approaches take advantage of the fact that the optimization problem regarding $|\I|$ items can be decomposed into $|\I|$ independent (and hence embarrassingly parallel)  optimization problems involving  $|\I|-1$ items each,  due to the identity
$
||\Y-\X\W||_F^2 = \sum_{j\in \I} | \Y_{.,j} -\X\W_{.,j}  |_2^2
$
. If each of the  $|\I|$ independent problems is solved based on an item-item matrix,
the total computational cost is hence  $\mathcal{O}(|\I|(|\I|-1)^{2.376})$. Note that the  computational cost of solving those  $|\I|$ problems is larger by almost a factor of $|\I|$ than training \sae{}, which requires only a  \emph{single} regression problem to be solved. In practice, however, \slim{}  and its variants are trained on the user-item-interactions in  $\X$, which may incur additional computational cost.
This explains the vastly reduced training-times of \sae{}   observed in our experiments in Section \ref{sec_exp}.

\begin{algorithm}[t]
\SetAlgoNoLine
\KwIn{data Gram-matrix $G := X^\top X \in \mathbb{R}^{|\I|\times|\I|}$, \\
\hspace{9 mm}  L2-norm regularization-parameter $\lambda\in \mathbb{R}^+$.}
\KwOut{weight-matrix $B$ with zero diagonal (see Eq. \ref{eq_w2}).}
$diagIndices$ = numpy.diag\_indices($G$.shape[0])  \\
$G[diagIndices]$ += $\lambda$  \\
$P$ = numpy.linalg.inv($G$) \\
$B$ =  $P$ / (-numpy.diag($P$)) \\
$B[diagIndices]$ = 0  \\
\caption{Training in Python 2 using numpy}
\label{algo}
\end{algorithm}

In practice, the wall-clock time depends  crucially on the fact if the  number of items $|\I|$ is sufficiently small such that the weight matrix fits into memory, so that the matrix inversion can be computed in memory. This was the case in our experiments in Section \ref{sec_exp}.

\section{Related Work}
\label{sec_related}

\sae{}  can be viewed as an autoencoder, as a modified version of \slim{}, and  a neighborhood-based approach. We discuss each of the three related approaches in the following.

\subsection{Deep Learning and Autoencoders}
While the area of collaborative filtering has long  been dominated by matrix factorization approaches, recent years have witnessed a surge in deep learning approaches \cite{liang18,sedhain15, zheng16,wu16,he17, hidasi15,hidasi17, cheng16}, spurred by their great successes in other fields. Autoencoders provide the model architecture that fits exactly the (plain-vanilla) collaborative filtering problem. While various network architectures have been explored, it was found that deep models with a large number of hidden layers typically do not obtain a notable improvement in  ranking accuracy in collaborative filtering, compared to 'deep' models with only one, two or three hidden layers, e.g.,  \cite{sedhain15, zheng16,he17,  liang18}, which is in stark contrast to other areas, like computer vision. A combination of deep and shallow elements in a single model was proposed in \cite{cheng16}.

In contrast, \sae{}  has no hidden layer. Instead, the self-similarity of each item in the input and output layer is constrained to zero, see also Figure \ref{fig_sae}. As a consequence, the model is forced to learn the similarity of an item in the output layer in terms of the \emph{other} items in the input layer. The surprisingly good empirical results of \sae{}  in Section \ref{sec_exp} suggest that this constraint might be more effective than using hidden layers with limited capacity as to force the model to generalize well to unseen data.

\subsection{\slim{}  and Variants}
While the \slim{} model \cite{ning11} has shown competitive empirical results in numerous papers, it is  computationally expensive to train, e.g., see  \cite{ning11,liang18} and Section \ref{sec_complexity}. This has sparked follow-up work proposing various modifications. In \cite{levy13}, both constraints on the weight matrix (non-negativity and zero diagonal) were dropped, resulting in a regression problem with an elastic-net regularization.   While competitive ranking results were obtained  in \cite{levy13}, in the experiments in \cite{liang18} it was found that its performance was considerably below par.
The square loss in \slim{} was replaced by the logistic loss in \cite{sedhain16}, which entailed that both constraints on the weight matrix could be dropped, as argued by the authors. Moreover,  the L1-norm regularization was dropped, and a user-user weight-matrix was learned instead of an item-item  matrix. 

All these approaches take advantage of the fact that the optimization problem decomposes into  independent and embarrassingly parallel  problems. As discussed in the previous section, however,  this is several orders of magnitudes  more costly than training \sae{}  if the weight matrix fits into memory.

Most importantly, while  those modifications of \slim{} dropped the constraint of a zero diagonal in the weight matrix, it is retained  in \sae. In fact, we found it to be the most crucial property for achieving improved ranking accuracy (see Section \ref{sec_exp}). As we showed in Section \ref{sec_closedform}, this constraint can be easily included into the training objective via the method of Lagrangian multipliers, allowing for a closed-form solution.

Compared to  \slim{}  \cite{ning11}, we dropped the constraint of non-negative weights, which we found to greatly improve ranking accuracy in our experiments (see Table \ref{tab_dawen} and Figure \ref{fig_weights}). Moreover, we  did not use L1-norm regularization for computational efficiency. We also did not find sparsity to noticeably improve ranking accuracy (see Section \ref{sec_exp}).  The learned weight matrix $\Wzero$ of \sae{} is dense.

Also note that extensions to \slim{} like cSLIM \cite{ning12}, can be turned into an analogous extension of \sae.

\subsection {Neighborhood-based Approaches}
\label{sec_nn}
Numerous neighborhood-based approaches have been proposed in the literature (e.g., see \cite{koen14, volkovs15} and references therein). While model-based approaches were found to achieve better ranking accuracy on some data sets, neighborhood-based approaches dominated on others, e.g.,  the Million Song Data Competition on Kaggle \cite{mcfee12,aiolli13}. Essentially, the co-occurrence matrix (or some modified variant) is typically used as item-item or user-user similarity matrix in neighborhood-based methods. These approaches are usually heuristics, as the similarity matrix is  not learned by optimizing an objective function (loss function or likelihood). More importantly, the closed-form solution derived in Eqs. \ref{eq_pp} and \ref{eq_w2} reveals that the \emph{inverse} of the data Gram  matrix is the conceptually  correct similarity matrix,\footnote{In fact,  inverse matrices are used in many areas, for instance, the inverse covariance matrix in the Gaussian  density function or in the Mahalanobis distance.} see Section \ref{sec_interpretation} for more details.  This is in contrast to the typical neighborhood-based approaches, which use the data Gram-matrix without inversion.
The use of the conceptually correct, inverse matrix in \sae{}  may explain the improvement observed in Table \ref{tab_nn} compared to the heuristics used by state-of-the-art  neighborhood approaches.

\section{Experiments}
\label{sec_exp}
In this section, the proposed \sae{}  model is empirically compared to several state-of-the-art approaches, based on two papers that provided publicly available code for reproducibility of results \cite{liang18,volkovs15}. Both papers together cover linear, non-linear, deep and probabilistic models, as well as neighborhood-based approaches.
\subsection{Experimental Set-up}
We will only summarize the experimental set-ups used in these papers, and refer the reader to these papers for  details.

{\bf Summary of  Set-up in \cite{liang18}:} 
This paper considers the following models:
\begin{itemize}

\item Sparse Linear Method  (\slim{}) \cite{ning11}. Besides the original model, also a   computationally faster approximation  (which drops the constraints on the weights) \cite{levy13} was considered, but its results were not  found  to be on par with the other models in the experiments in \cite{liang18}.

\item Weighted Matrix Factorization (\wmf) \cite{hu08,pan08}, a linear model with a latent representation of users and items.

\item Collaborative Denoising Autoencoder (\cdae) \cite{wu16}, a non-linear model with one hidden layer.

\item denoising autoencoder (\mdae) and variational autoencoder (\mvae) \cite{liang18}, both trained using the multinomial likelihood, which was found to outperform the Gaussian and logistic likelihoods. Best results were obtained in \cite{liang18} for the \mvae{}  and \mdae{} models that were rather shallow 'deep models', namely with a 200-dimensional latent representation, as well as a 600-dimensional hidden layer in both the encoder and decoder. Both models are non-linear, and \mvae{}  is also probabilistic.

\end{itemize}
Three data sets were used in the experiments in \cite{liang18}, and were pre-processed and filtered for items and users with a certain activity level, resulting in the following data-set sizes, see \cite{liang18} for details:\footnote{The code regarding  \emph{ML-20M} in \cite{liang18} is publicly available at {\tt https://github.com/dawenl/vae\_cf}. The authors kindly provided the code for the other two  data sets upon request.}
\begin{itemize}
\item MovieLens 20 Million (\emph{ML-20M}) data \cite{movielens20mio}: 136,677 users and 20,108 movies with about 10 million interactions,
\item Netflix Prize (\emph{Netflix}) data \cite{netflixdata}: 463,435 users and 17,769 movies with about 57 million interactions,
\item Million Song Data (\emph{MSD}) \cite{msddata}: 571,355 users and 41,140 songs with about 34 million interactions.
\end{itemize}
The evaluation in \cite{liang18} was conducted in terms of \emph{strong generalization}, i.e., the training, validation and test sets are disjoint in terms of users. This is in contrast to \emph{weak generalization}, where the training and test sets are disjoint in terms of the user-item interaction-pairs, but not in terms of users. Concerning  evaluation in terms of ranking metrics, Recall@$k$ for $k\in\{20,50\}$ as well as Normalized Discounted Cumulative Gain, NDCG@100 were used  in \cite{liang18}.

{\bf Summary of Set-up in \cite{volkovs15}:} Their paper focuses on neighbor-hood-based approaches, and
the authors publicly shared code\footnote{{\tt http://www.cs.toronto.edu/$\sim$mvolkovs/SVD\_CF.zip}} regarding the experiments in their table 2 in \cite{volkovs15}, albeit only for the data-split where the training data  was comprised of (at most) 30\% of each user's interactions (and the remainder was assigned to the  test data), which restricted our experimental comparison to this single split in  Table \ref{tab_nn}.
 They used the  MovieLens 10 Million (\emph{ML-10M}) data \cite{movielens20mio}, which was binarized in \cite{volkovs15} and is comprised of 
69,878 users and 
10,677 movies with 10 million interactions.  Their evaluation was done in terms of \emph{weak generalization}, and NDCG@10  was used as ranking metric for evaluation in \cite{volkovs15}.

\begin{figure}[t]
\begin{center}
\includegraphics[height=4cm]{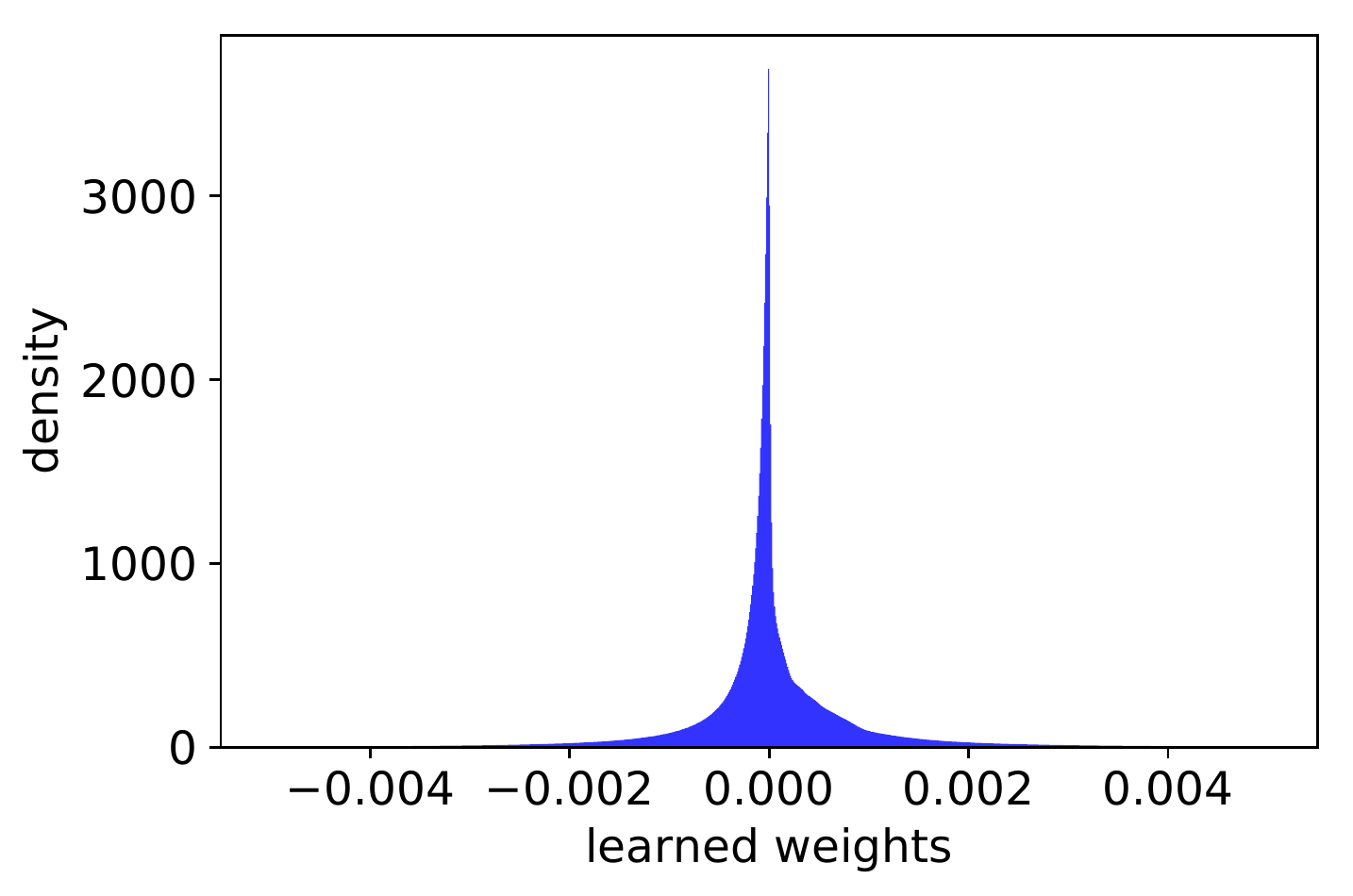}
\end{center}
\caption{Histogram of the weights learned on \emph{Netflix} data.}
\label{fig_weights}
\end{figure}

\subsection{Results}
Despite the    simplicity of \sae{}, we observed that \sae{}  obtained considerably better ranking accuracy than any of the competing models on most of the  data sets. This remarkable empirical result is discussed in detail in the following.

{\bf Comparison to} \slim{}:  Table \ref{tab_dawen} shows that \sae{}  achieved notably increased accuracy compared to \slim{} on all the data sets. This suggests that dropping the L1-norm regularization as well as the non-negativity constraint on the learned weights is beneficial. Our analysis indicates that the latter is especially important: as illustrated in Figure \ref{fig_weights} on the \emph{Netflix} data (the histograms for \emph{ML-20M} and \emph{MSD} data look almost identical up to re-scaling, and are omitted), the learned weights in \sae{}  are distributed around 0. Interestingly, it turns out that about 60\% of the learned weights are negative on all the data sets in our experiments (regarding both papers \cite{liang18,volkovs15}). This indicates that it is crucial  to learn also the dissimilarity (negative weights) between items besides their similarity (positive weights).
Moreover, when we simply set the negative weights to zero (see  \sae{}$\ge 0$ in Table \ref{tab_dawen}), which obviously is not the optimal non-negative solution, the resulting accuracy drops and is very close to the one of \slim{}.
Apart from that, note that \sae{}$\ge 0$ is still quite dense (40\% positive weights) compared to \slim{}, which indirectly indicates that the sparsity of \slim{} (due to L$_1$-norm regularization) did not noticeably improve the ranking accuracy of \slim{} in our experiments.

Regarding regularization,  the optimal L2-norm regularization parameter ($\lambda$) for \sae{}   is about 500 on \emph{ML-20M}, 1,000 on \emph{Netflix}, and 200 on \emph{MSD}. These values are much larger than the typical values used for \slim{}, which often are of the order of 1, see \cite{ning11}. Note that \slim{} additionally uses L1-norm regularization, and hence has much fewer (non-zero) parameters than \sae.  

As expected based on Section \ref{sec_complexity}, we also found the (wall-clock) training-time of \sae{}  to be smaller by several orders of magnitude compared to \slim{} : \cite{liang18} reports that parallelized grid search for \slim{} took about two weeks on the \emph{Netflix} data, and the \emph{MSD} data was 'too large for it to finish in a reasonable amount of time' \cite{liang18}. In contrast,  training \sae{}  on the \emph{Netflix} data took less than two minutes, and on the  \emph{MSD} data less than 20 minutes on an AWS instance with 64 GB RAM and 16 vCPUs in our experiments. These times have to be multiplied by the number of different hyperparameter-values to be grid-searched.  Note, however, that \sae{}  only has a single hyperparameter (regarding L2-norm regularization), while \slim{} has two hyperparameters (concerning L1 and L2 norms) to be jointly optimized.

\begin{table}
\caption{Ranking accuracy (with standard errors of about 0.002, 0.001, and 0.001 on the \emph{ML-20M}, \emph{Netflix}, and \emph{MSD} data, respectively), following the experimental set-up in \cite{liang18}.}
\label{tab_dawen}
\begin{tabular}{lrrr}
\hline
{\bf(a)\emph{ ML-20M}}  &   Recall@20    &    Recall@50    &    NDCG@100\\
popularity    &    0.162    &    0.235    &   0.191 \\
\sae{}     &    0.391    &    0.521    &   0.420 \\
\sae{}$\ge 0$    &  0.373    &  0.499  &  0.402\\
\multicolumn{4}{l}{ results reproduced from \cite{liang18}:}\\
\slim{}    &    0.370    &    0.495    &    0.401\\
\wmf{}     &    0.360    &    0.498    &    0.386\\
\cdae{}     &    0.391    &    0.523    &    0.418\\
\mvae{}     &    0.395    &    0.537    &    0.426\\
\mdae{}     &    0.387    &    0.524    &    0.419\\
\hline
{\bf (b) \emph{Netflix}}&& &\\
popularity    &    0.116    &    0.175    &   0.159 \\
\sae{}     &    0.362    &    0.445    &   0.393 \\
\sae{}$\ge 0$    &  0.345    &  0.424  &  0.373\\
\multicolumn{4}{l}{ results reproduced from \cite{liang18}:}\\
\slim{}    &    0.347    &    0.428    &    0.379\\
\wmf{}     &    0.316    &    0.404    &    0.351\\
\cdae{}     &    0.343    &    0.428    &    0.376\\
\mvae{}     &    0.351    &    0.444    &    0.386\\
\mdae{}     &    0.344    &    0.438    &    0.380\\
\hline
{\bf (c)     \emph{MSD}}& & &\\
popularity    &    0.043    &    0.068    &   0.058 \\
\sae{}     &    0.333    &    0.428    &   0.389 \\
\sae{}$\ge 0$    &  0.324    &  0.418  &  0.379\\
\multicolumn{4}{l}{ results reproduced from \cite{liang18}: }\\
\slim{}    &    \multicolumn{3}{c}{--- did not finish in \cite{liang18} ---}     \\
\wmf{}     &    0.211    &    0.312    &    0.257\\
\cdae{}     &    0.188    &    0.283    &    0.237\\
\mvae{}     &    0.266    &    0.364    &    0.316\\
\mdae{}     &    0.266    &    0.363    &    0.313\\
\hline
\end{tabular}
\end{table}

\begin{figure}[t]
\begin{center}
\includegraphics[height=4cm]{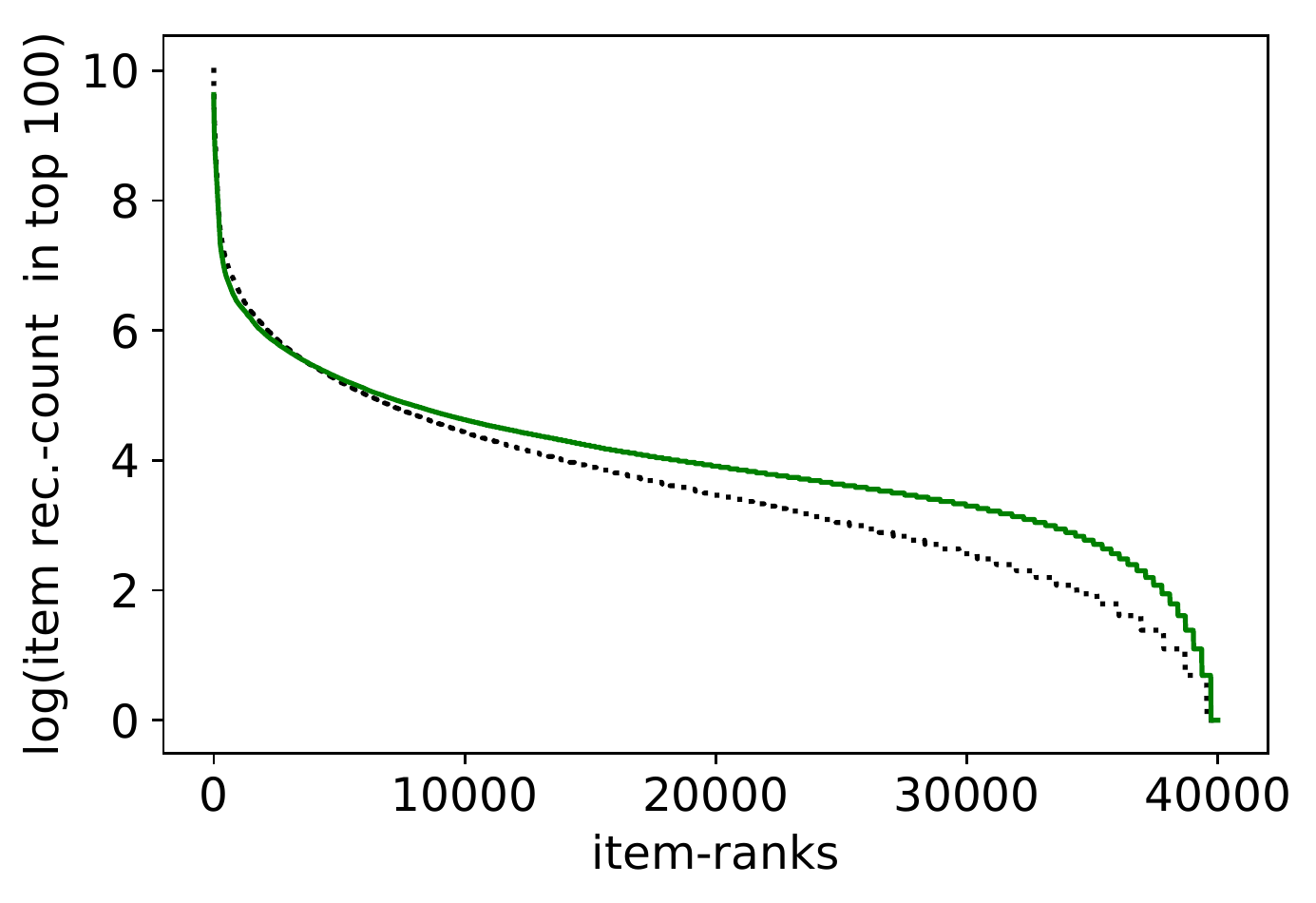}
\end{center}
\caption{\sae{} (green) recommends  long-tail items more often in the top-100, compared to \mvae{} (dotted), on \emph{MSD} data.}
\label{fig_recs}
\end{figure}

{\bf Comparison to linear and deep non-linear models in \cite{liang18}:}
Table \ref{tab_dawen} shows that \sae{}  was consistently outperformed on only the \emph{ML-20M} data, and only by a small margin by the best competing model (\mvae).  On the  \emph{Netflix} and \emph{MSD} data, \sae{}  obtained significantly better ranking results than any of the competing linear, non-linear, deep or probabilistic models evaluated in \cite{liang18}. On the \emph{MSD} data, \sae{}  even improved over the best competing model by  25\%, 17\% and 23\% regarding Recall@20, Recall@50, and NDCG@100, respectively. This is consistent with the results of the Million Song Data Challenge on Kaggle \cite{mcfee12}, where neighborhood-based approaches were found to vastly outperform model-based approaches \cite{aiolli13}. As discussed in Section \ref{sec_nn}, \sae{}  may also be viewed as a principled neighborhood approach.

As to explain \sae's relative improvements from \emph{ML-20M} via  \emph{Netflix} to \emph{MSD} data, various properties of the data sets may be considered. As shown by table 1 in \cite{liang18}, the number of user-item interactions, and the sparsity of the data sets  do not appear well correlated  with \sae's relative performance in Table \ref{tab_dawen}. Only the number of users correlates well with the improvements of \sae{}  over the competing models, which, however, appears to be spurious.

The explanation can be understood in terms of the tradeoff between recommending generally popular items vs. personally relevant items to each user, which is supported by two empirical findings: (1) we evaluated the popularity model in Table \ref{tab_dawen} as an additional baseline, 
where  the items are ranked by their popularities (i.e.,  the number of users who interacted with an item). These unpersonalized recommendations obviously ignore the personalized relevance to a user. Table \ref{tab_dawen} shows that this  popularity model obtains better accuracy on the \emph{ML-20M} data than it does on the \emph{Netflix} data, while its accuracy is considerably reduced on the \emph{MSD} data. This suggests that good recommendations on the  \emph{MSD} data have to focus much more on personally relevant items rather than on generally popular items, compared to the  \emph{ML-20M} and \emph{Netflix} data. (2) When counting how often an item was recommended in the top-100 across all test-users, and then ranking the items by their counts, we obtained Figure \ref{fig_recs} for the \emph{MSD} data: it shows that \sae{} recommended long-tail items more often than \mvae{} did. In contrast, there was almost no difference  between the two approaches on either of the data sets \emph{ML-20M} and \emph{Netflix} (figures omitted due to page limit).

The notable  improvement of \sae{}  over the other models on the   \emph{MSD} data suggests that it is able to better recommend personally relevant items on this data set. On the other hand, \sae{}'s results on  the  \emph{ML-20M} and \emph{Netflix} data suggest that it is also able to make recommendations with an increased focus on popular items. We suspect that \sae's large number of parameters, combined with its constraint regarding self-similarity of items,  provides it with sufficient flexibility to adapt to the various data sets. In contrast, the model architectures based on hidden layers with limited capacity seem to be unable to adapt well to the increased degree of  personalized relevance in the \emph{MSD} data.

\begin{table}[t]
\caption{Comparison to the neighborhood-approaches in \cite{volkovs15}: \sae{} considerably improves over 'ii-SVD-500' \cite{volkovs15}.}
\label{tab_nn}
\begin{tabular}{llllll}
\hline
        &       \sae{}          &     \sae{}              & \multicolumn{3}{l}{reproduced from \cite{volkovs15}: }\\
       &        & $\,\,\,\ge 0$   &   ii-SVD-500   & item-item    & \wmf{}  \\
\hline
NDCG@10   &      0.6258      &  0.6199       &       0.6113         &      0.5957   &    0.5969\\
\hline
\end{tabular}
\end{table}

{\bf Comparison to neighborhood-based approaches in \cite{volkovs15}:}
Considerable improvements were obtained in \cite{volkovs15} by first predicting the scores for all user-item interactions with a neighborhood-based approach ('item-item' in Table \ref{tab_nn}) that was followed by a low-rank singular value decomposition  of the predicted user-item score-matrix ('ii-SVD-500' in Table \ref{tab_nn}): an increase in NDCG@10 by 0.0156  and 0.0144 compared to  the baseline models  'item-item' and \wmf{}, respectively, as reproduced in Table \ref{tab_nn}. In comparison, \sae{} obtained  increases of 0.0301 and  0.0289  over the baseline models  'item-item' and \wmf{}, respectively, see Table \ref{tab_nn}. This is about twice as large an improvement as was obtained by the approach 'ii-SVD-500' proposed in \cite{volkovs15}.

Given the small size of this training data-set, a  large L2-norm regularization was required ($\lambda=3000$) for \sae{}. Like in the previous experiments, also here about 60\% of the learned weights were negative in \sae, and setting them to zero (\sae{}$\ge 0$ in Table \ref{tab_nn}) resulted in  a small decrease in NDCG@10, as expected. 
 In terms of wall-clock time, we found that training \sae{}  was about three times faster than computing merely the  factorization step in the ii-SVD-500 approach.

\section{Conclusions}
We presented a simple yet effective linear model for collaborative filtering, which combines the strengths of autoencoders and neighborhood-based approaches.
 Besides enabling efficient training (with savings of up to several orders of magnitude  if the model fits into memory), the derived closed-form solution also shows that the conceptually correct similarity-matrix to be used in neighborhood approaches is based on the \emph{inverse} of the given data Gram-matrix. In contrast, state-of-the-art neighborhood approaches typically use the data Gram-matrix directly. In our experiments, we found that allowing the learned weights to also take on negative values (and hence learn dissimilarities between items, besides similarities), was essential for the obtained ranking accuracy. Interestingly, the achieved ranking accuracy in our experiments  was on par or even (notably) better than those of various state-of-the-art approaches, including deep non-linear models as well as neighborhood-based approaches. This suggests that models where the self-similarity of items is constrained to zero may be more effective on sparse data than model architectures based on hidden layers with limited capacity. We presented a basic version as to illustrate the essence of the idea, and leave various modifications  and extensions for future work. Several practical extensions are outlined in \cite{steck19b}.

\begin{acks}
 I am very grateful to Tony Jebara and  Nikos  Vlassis for useful comments on an earlier draft, and especially to  Dawen Liang for suggestions on related work with publicly available code.
\end{acks}

\bibliographystyle{ACM-Reference-Format}

\end{document}